\documentclass[10pt,twoside]{epnt1p}
\usepackage{epsfig}

\usepackage{amssymb}
\usepackage{amsmath}

\usepackage{url}

\begin{document}

\begin{frontmatter}


\title{Discovery and Upper Limits in Search for Exotic Physics with Neutrino Telescopes}


\author[address1]{J. Conrad}

\address[address1]{KTH Stockholm, Fysik, AlbaNova University Centre, SE-10691 Stockholm}

\begin{abstract}
This note gives a short review of the statistical issues concerning upper limit calculation and claiming of discovery arising in the search for exotic physics with neutrino telescopes. Low sample sizes and significant instrumental uncertainties require special consideration. Methods for treating instrumental or theoretical uncertainties in the calculation of limits or discovery are described. Software implementing these methods is presented. The issue of optimization of analysis cuts and definition of sensitivity is briefly discussed.
\end{abstract}



\end{frontmatter}

\section{\label{sec:intro} Introduction}
Searches for exotic physics in neutrino telescopes pose challenges to statistical inference. In neutrino telescopes often physicists have to deal with low event counts and significant instrumental uncertainties which need to be taken into account. The small sample size makes the use of asymptotic methods doubtful. Presence of significant instrumental or theoretical uncertainties introduce nuisance parameters which need to be considered in the estimation of physics parameter intervals or in claims for discovery.\\
Neutrino astronomy has so far been in an infant state, the main aim of existing experiments being proof of principle with little hope of detecting neutrinos of cosmic and/or exotic origin. Analyses have therefore often been optimized for setting the most stringent upper limit. With the advent of the IceCube detector this situation might change, in the sense that discovery should become more likely. Analysis optimization methods should therefore be reviewed.\\
In the next section, we remind the reader of methods to calculate confidence intervals and to claim discovery. In section \ref{sec:nuisance} we present two methods for including uncertainties, followed by a discussion on analysis optimization.\\
For simplicity, unless otherwise stated,  throughout the paper we will consider a statistical process with probability density function (PDF) $P(n|s)$, i.e. the probability to observe $n$ given the parameter $s$.

\section{\label{sec:discovery} Claiming discovery and calculating confidence intervals}
Though often mixed up \cite{Cranmer:2005,Conrad:2006}, claiming discovery and calculation of confidence intervals are generally presented as two different cases in statistical literature, see e.g. \cite{Eadie:1971} for an exhaustive introduction. 
\paragraph*{Claiming discovery}
Claiming discovery is an example of {\it hypothesis testing}. In the search of exotic physics the null hypothesis, $H_{0}$, is usually ``No signal present'' and the alternative hypothesis, $H_{1}$, is ``Signal of exotic physics present''.  Discovery is claimed if the measurement result would be very unlikely under the condition that the null hypothesis is true. This is quantified by calculating the {\it p-value}:
\begin{equation}
p = P(T \geq T_{obs} | H_{0})
\end{equation}
where $T$ denotes a {\it test statistics}, which is a function of the observation and the involved hypotheses (see below), and $P$ is a probability density function describing the distribution of the test statistics under the null hypothesis.
If the p-value is smaller than some predetermined threshold, $\alpha_{sign}$, then the null hypothesis is rejected. A particular common choice of the threshold is the probability corresponding to 5$\sigma$ in an equivalent Gaussian distribution, i.e.  $\alpha_{sign} \sim 3 \cdot 10^{-7}$.\\
The test statistics which in a statistical sense is most optimal in distinguishing two alternative hypotheses is according to the Neyman-Pearson lemma the likelihood ratio:
\begin{equation}
T = \frac{\mathcal{L}(n|H_{0})}{\mathcal{L}(n|H_{1})}
\end{equation}
One useful property of the likelihood ratio is that under the assumption that the null hypothesis is true, asymptotically the following statement holds:
\begin{equation}
-2 \ln{T} \sim \chi^2  
\end{equation}
i.e. minus twice the logarithm of the likelihood ratio follows a $\chi^2$ distribution. Though this property is often used in particle physics it is not always true. Except for normality, there are a series of other requirements, some of which are violated even for quite common cases \cite{Conrad:2006}\cite{Demortier:2006}. An experimenter intending to use the $\chi^2$ distribution should therefore be convinced that this is justified and in case of doubt, calculate the distribution of the test statistics under the null hypothesis for example by  using Monte Carlo  simulations.\\
The threshold probability in order to decide if the null hypothesis should be rejected is based on the number of false detection one is willing to accept and on the number of searches which are performed. For example, in the context of one of the large CERN experiments, there might be about 50000 independent channels \cite{Feldman:2005}, i.e. 500 channels, 1000 resolution elements each. This implies a false positive detection rate of about 1.5 \% if one requires a 5$\sigma$ detection. It is not obvious, that a similar reasoning would lead to the same requirement in case of searches with neutrino telescopes.

\paragraph*{Calculation of confidence intervals}
Consider $s$  a fixed (unknown) parameter. In frequentist statistics\footnote{the Bayesian calculation is somewhat simpler, since here the interval can be found by integrating in the probability density function $P(s|n)$, where the parameter is considered a random variable}, one defines a confidence interval [$s_1,s_2$] as a member of a set of intervals for which:
\begin{equation}
P(s \in [s_1,s_2]) = 1-\alpha \,\,\,\,\,\,,  \,\,\,\, for  \,\,\,\,\, all \,\,\,s
\label{eq:cov}
\end{equation}
These sets are to be constructed using only $P(n|s)$, $s$ being a fixed but unknown parameter and not $P(s|n)$. The latter would make the method Bayesian, since $s$ is treated as a random variable. The frequentist construction has been introduced by Neyman \cite{Neyman:1937a}. A special case, which is very common in high energy physics has been proposed by Feldman \& Cousins \cite{Feldman:1998a}. Other common approximative methods exist which make use only of the likelihood function, e.g. {\cite{Eadie:1971,Barlow:2003}.\\
\noindent
All frequentist methods which fulfill the condition in equation \ref{eq:cov} are said to have {\it coverage}\footnote{To be more precise, in frequentist statistics in addition the probability has to be defined in terms of repeated identical experiments.}. Methods which do not have coverage are not valid or need correction\footnote{In general, physicists tend to accept over-coverage (meaning, being conservative), whereas under-coverage is considered unacceptable.}. In case of doubt, coverage has to be demonstrated for example by using Monte Carlo simulations.

\section{\label{sec:nuisance} Treatment of nuisance parameters}

{\it Nuisance parameters } are parameters which enter the data model, but which are not of prime interest. The probably most common example is the expected background in a Poisson process. In a usual physics experiment, only confidence intervals on the parameter of primary interest (for example signal flux or cross-section) are of interest. Thus, ways have to be found to marginalize the nuisance parameter. There are two common approaches: \\

\noindent 
In the first method, the PDF without uncertainty in nuisance parameters is replaced by one where there is an integration over all possible true values of the nuisance parameter {\it(integration method)}:
\begin{equation}
P(n|s,b_{true}) \longrightarrow \int_0^{\infty}{P(n|s,b_{true}) P(b_{true}|b_{est}) d\,b_{true}}
\label{eq:integration}
\end{equation}
Here $b_{true}$ is the true value of the nuisance parameter and $b_{est}$ is its estimate. Since the integrated PDF is describing the probability of the true value given its estimate (and not vice versa) this method is Bayesian. Some prior probability distribution of the true value of the nuisance parameter has to be assumed.\\
\noindent
In the other common method, the PDF is replaced by one where for each $s$ the PDF is maximized with respect to the nuisance parameters {\it(profiling method)}
\begin{equation}
P(n|s,b_{true}) \longrightarrow \max_{b_{true}}{\mathcal{L}(n|s,b_{true})}
\label{eq:profile}
\end{equation}
with notation as above. This method is completely frequentist, since it never treats $b_{true}$ as a random variable. Therefore the argument of the maximisation is a likelihood function and not a PDF.}\\

\noindent
Both approaches have recently been subjected to detailed studies regarding their coverage, i.e. with applications to confidence intervals \cite{Tegenfeldt:2004dk,Rolke:2004mj,Conrad:2005b}. For typical problems arising in neutrino telescopes they perform satisfactory. In the context of the LHC searches there are indications the Bayesian method under-covers badly, whereas the profiling method still seems to work fine \cite{Cranmer:2005}.\\
Software has been developed which implements the  {\it integration} and {\it profiling} method for typical problems arising in neutrino telescope analyses.\\

\noindent
\texttt{pole++} is a C++ library of classes which was developed based on the method presented in \cite{Conrad:2002a}. It allows calculating confidence intervals using the Feldman \& Cousins method with integration of nuisance parameters and coverage studies. The signal process considered is a Poisson with known (possibly uncertain) background with different models for nuisance parameters. Several experiments with correlated or uncorrelated uncertainties in the nuisance parameters can be combined. The pole++ library can be obtained from http://cern.ch/tegen/statistics.html \\
\texttt{TRolke} is class which is part of the \texttt{ROOT} analysis package \cite{ROOT}.  It treats a Poisson signal with background process with seven different models for nuisance parameters, which are marginalized using the profiling method. For extraction of the confidence intervals the likelihood function is used. For a description of the method see \cite{Rolke:2004mj}.\\
\noindent
It should be noted that the \texttt{MINUIT} package \cite{James:1975dr} (in particular using the \texttt{MINOS} facility) also applies the profiling method and is completely general as long as the likelihood function can be written down. For the special case of a Poisson process \texttt{TRolke} applies some improvements.\\

\noindent
Though profiling and integration are here presented in the context of confidence intervals, both approaches can also be used in hypothesis testing. The substitutions in equations \ref{eq:integration} and \ref{eq:profile} then have to be applied to the likelihoods. The statistical properties to be studied in this case would be the distribution of the test statistics under the null hypothesis as well as possibly the power function (see next section).


\section{\label{sec:optimization} Analysis optimization}
Analyses are optimized defining some {\it figure of merit} (FOM), which will be maximized (or minimized) with respect to some cut value $t$\\
In searches with neutrino telescopes, it is often chosen to try to set the most stringent upper limit leading to the introduction of Model Rejection Factor \cite{Hill:2002nv}: 
\begin{equation}
MRF = \frac{<s_{1-\alpha}>}{n_s}
\end{equation}
Here, $s_{1-\alpha}$ denotes the (1-$\alpha$) confidence level upper limit on $s$. The mean is taken over the Poisson distribution with no signal. In case of presence of uncertainties in nuisance parameters the corresponding mean becomes:
\begin{equation}
<s_{1-\alpha}> = \int_0^{\infty} \int_0^{\infty}\,dn \, db_{est} s_{1-\alpha} P(n|b_{est}) P(b_{est}|b_{true}) 
\end{equation}
Unless the size of the uncertainties depends on the cut value or assymetric PDFs for the nuisance parameters have to be considered, the mean calculated using only the Poisson distribution should be sufficient for the calculation of the FOM.

\noindent
In case a decision on whether to set a limit or claim discovery beforehand is not desired, the figure of merit will have to be based on a sensitivity region which is meaningful both if the experimenter would report a limit and if the experimenter wants to claim discovery. Punzi \cite{Punzi:2003} suggests to define the sensitivity region by:
\begin{equation}
1-\beta_{\alpha_{sign}}(s) > 1-\alpha_{CL}
\label{eq:Punzi}
\end{equation}
Where $\alpha_{sign}$ is the significance which the experimenter requires to claim discovery and $\alpha_{CL}$ is the confidence level required in case the experimenter wants to calculate a limit. $\beta (s)$ is the so called power function. Power is a concept arising in hypothesis testing. The power of a test is the probability to reject the null-hypothesis given the alternative hypothesis is true. The concept can also be applied to confidence intervals \cite{Conrad:2005b}. This definition of sensitivity fulfills several desirable properties: for example if $s$ is inside the region defined by equation \ref{eq:Punzi}, then there is a probability of at least $1-\alpha_{CL}$ that it will be discovered. From the sensitivity a FOM can be calculated. Simple expressions of the FOM for common problems can be found in \cite{Punzi:2003}.

\section{Conclusions}
Major statistical challenges to be faced in searches for new physics with neutrino telescopes are low statistics, large instrumental uncertainties and the desire to detect an unknown process while at the same time being able to set stringent upper limits.\\
In this note we point out and briefly discuss different solutions to the above challenges. There are several methods existing to calculate confidence intervals in the presence of instrumental uncertainties which have been tested and behave well also for the small sample sizes and relatively large uncertainties usually encountered in neutrino astronomy. Code performing the necessary calculations is readily available. Sensitivity regions linking the power of a hypothesis test with the confidence level of a confidence interval yield figure of merits which can be used to optimize analysis with respect to both discovery and stringent limits.

\section*{Acknowledgments}

{\small 
I thank the organizers of the conference for the invitation to give this talk and Olga Botner and Allan Hallgren for reading the manuscript.

}

\setcounter{section}{0}
\setcounter{subsection}{0}
\setcounter{figure}{0}
\setcounter{table}{0}
\newpage
\end{document}